\documentclass[twocolumn]{autart}    

\usepackage{graphicx}          
\usepackage{float}
\usepackage[colorlinks,linkcolor=blue]{hyperref}
\usepackage[latin1]{inputenc}
\usepackage{graphicx}
\usepackage{amssymb}
\usepackage{verbatim}
\usepackage{epsfig}
\usepackage[labelfont=bf]{caption}
\usepackage{enumerate}
\usepackage{subfig}
\usepackage{epstopdf}

\usepackage{stfloats}
\usepackage{amssymb}

\usepackage{graphicx}
\usepackage{bm}
\usepackage{color}
\usepackage{multirow}
\usepackage{mathrsfs}
\usepackage{lineno,hyperref}
\usepackage{lipsum}
\usepackage{stfloats}
\usepackage{titletoc}
\usepackage{appendix}
\usepackage{dsfont}
\usepackage{mathtools}

\def\begquo{\begin{quote}}
	\def\endquo{\end{quote}}
\def\begequarr{\begin{eqnarray}}
\def\endequarr{\end{eqnarray}}
\def\begequarrs{\begin{eqnarray*}}
	\def\endequarrs{\end{eqnarray*}}
\def\begarr{\begin{array}}
	\def\endarr{\end{array}}
\def\begequ{\begin{equation}}
\def\endequ{\end{equation}}
\def\lab{\label}
\def\begdes{\begin{description}}
	\def\enddes{\end{description}}
\def\begenu{\begin{enumerate}}
	\def\begite{\begin{itemize}}
		\def\endite{\end{itemize}}
	\def\endenu{\end{enumerate}}

\def\lef[{\left[\begin{array}}
	\def\rig]{\end{array}\right]}

\def\begcen{\begin{center}}
	\def\endcen{\end{center}}
\def\begdef{\begin{definition}}
	\def\enddef{\end{definition}}
\def\begsubequ{\begin{subequations}}
	\def\endsubequ{\end{subequations}}

\def\begmat#1{\begin{bmatrix}#1\end{bmatrix}}
\def\begali#1{\begin{align}{#1}\end{align}}
\def\begalis#1{\begin{align*}{#1}\end{align*}}

\def\liminf{\lim_{t \to \infty}}

\def\L2e{{\cal L}_{2e}}

\def\rea{\mathbb{R}}

\usepackage{color}

\usepackage[prependcaption,colorinlistoftodos]{todonotes}

\begin{document}
\begin{frontmatter}

\title{State Observation of LTV Systems with Delayed Measurements: A Parameter Estimation-based Approach} 

\thanks[footnoteinfo]{This paper was not presented at any IFAC 
meeting. Corresponding author N.~Nikolaev. Tel. +79213090016.}

\author[1]{Alexey Bobtsov}\ead{bobtsov@mail.ru},    
\author[1]{Nikolay Nikolaev}\ead{nanikolaev@itmo.ru},           
\author[2]{Romeo Ortega}\ead{romeo.ortega@itam.mx},  
\author[3]{Denis Efimov}\ead{denis.efimov@inria.fr}.  

\address[1]{Department of Control Systems and Robotics, ITMO University, Kronverkskiy av. 49, Saint-Petersburg, 197101, Russia}  
\address[2]{Departamento Acad\'{e}mico de Sistemas Digitales, ITAM, Ciudad de M\'exico, M\'{e}xico}             
\address[3]{Inria, Univ. Lille, CNRS, UMR 9189 - CRIStAL, F-59000 Lille, France}        


\begin{abstract}                          
In this paper we address the problem of state observation of linear time-varying systems with delayed measurements, which has attracted the attention of many researchers---see \cite{SANGARKRS} and references therein. We show that, adopting the parameter estimation-based approach proposed in \cite{ORTetalaut20,ORTetalscl15}, we can provide a very simple solution to the problem with reduced prior knowledge.
\end{abstract}

\end{frontmatter}

\section{Main result}
\lab{sec1}
%
\begin{prop}\em
Consider a linear time-varying (LTV) system
\begali{
\nonumber
\dot{x}(t)&=A(t)x(t)+B(t)u(t)\\
\lab{sys}
y(t)&=C(\varphi(t))x(\varphi(t)),
}
for $t\geq0$ with $x(t) \in \rea^n$, $u(t) \in \rea^m$, $y(t) \in \rea^q$, where  $\varphi(t)$ is a {\em known} delay function verifying
$$
t \geq \varphi(t) \geq 0.
$$
The generalized parameter estimation-based observer
\begalis{
\dot{\xi}(t)&=A(t)\xi(t)+B(t)u(t)\\
\dot{\Phi}(t)&=A(t)\Phi(t),\;\Phi(0)=I_n\\
\hat x(t)&=\xi(t) - \Phi(t) \hat \theta(t),
}
with the gradient parameter estimator
\begali{
\lab{parest}
\nonumber
\dot {\hat \theta}(t)&=\Gamma \Phi^\top (\varphi(t))C^\top (\varphi(t))[C(\varphi(t))\xi(\varphi(t))\\
&-y(t)-C(\varphi(t)) \Phi(\varphi(t)  )\hat \theta(t)],
}
with $\Gamma>0$, which ensures
$$
\liminf |\hat x(t) -x(t)|=0,\quad (exp.)
$$
provided $C(t) \Phi(t)$ is persistently exciting (PE) \cite{SASBOD}, that is, there exists positive constants $T$ and $\delta$ such that
\begequ
\lab{pe}
\int_t^{t+T} C(s) \Phi(s)\Phi^\top(s)C^\top(s)ds \geq \delta I_q,\; \forall t \geq 0.
\endequ 
\end{prop}

\begin{pf}
Define the error signal $e(t):=\xi(t)-x(t)$, which satisfies
$$
\dot{e}(t)=A(t)e(t),
$$
hence 
$$
e(t)=\Phi(t)\theta,
$$
with $\theta :=e(0)$. Consequently,
\begequ
\lab{x}
x(t)=\xi(t) - \Phi(t) \theta.
\endequ

The output of the system \eqref{sys} then satisfies
$$
y(t)=C(\varphi(t))\left[\xi(\varphi(t))-\Phi(\varphi(t))\theta\right].
$$
From which we get the linear regression equation
$$
C(\varphi(t))\xi(\varphi(t))-y(t)=C(\varphi(t))	\Phi(\varphi(t))\theta,
$$
that, replacing in \eqref{parest}, yields the parameter error equation
$$
\dot {\tilde \theta}(t)=-\Gamma \Phi^\top (\varphi(t))C^\top (\varphi(t)) C(\varphi(t)) \Phi(\varphi(t)) \tilde \theta(t),
$$
with $\tilde \theta(t):=\hat \theta(t)- \theta$. 

Invoking standard adaptive control arguments \cite[Theorem 2.5.1]{SASBOD} we conclude that, the PE assumption \eqref{pe} ensures
$$
\liminf |\tilde \theta(t)|=0,\quad (exp.)
$$
The proof is completed noting that
$$
\hat x(t)-x(t)=-\Phi(t)\tilde \theta(t).
$$ 
\end{pf}

\begin{rem}{\bf{1}}
Another, more complex, solution to this problem that requires the {\em knowledge} of $\dot \varphi(t)$ is reported in \cite{SANGARKRS} under the classical assumption of existence of an exponentially stable Luenberger observer for the LTV system \eqref{sys} with $\varphi(t)=t$, {\em i.e.} \cite[Assumption 2]{SANGARKRS}. That estimator requires a more sophisticated implementation since it is based on a PDE representation of the delay, with an observer designed for the coupled LTV-PDE system. As is well known \cite{RUGbook} the PE assumption made here is equivalent to uniform complete observability of the pair $(C(t),A(t))$ and this, in its turn, is a {\em sufficient} condition for the verification of \cite[Assumption 2]{SANGARKRS}. 
\end{rem}

\begin{rem}{\bf{2}}
The PE assumption made here can be relaxed by the significantly weaker condition of {\em interval excitation} \cite{KRERIE} using the finite convergence time version of the dynamic regressor extension and mixing (DREM) estimator proposed in \cite{ORTetalaut19}, with the additional advantage of ensuring convergence in {\em finite time}. Adding fractional powers in the estimator, as done in \cite{WEAB,WEB}, it is also possible to achieve convergence in {\em fixed} time. The details are omitted for brevity.\end{rem}

\begin{rem}{\bf{3}}
Notice that if the state transition matrix converges to zero, {\em e.g.}, for a constant, Hurwitz matrix $A$, the estimation error converges to zero independently of the excitation conditions. In this case, the observer behaves like an open-loop emulator.
\end{rem}

\begin{rem}{\bf{4}}
Following \cite{SANGARKRS} assume that the function $\varphi(t)$ admits a
(piece-wise) continuous time derivative, then the PE condition can
be rewritten as follows: there exist $T>0$ and $\delta>0$ such that
\[
\int_{\varphi(t)}^{\varphi(t+T)}\dot{\varphi}^{-1}(s)\Phi^{\top}(s)C^{\top}(s)C(s)\Phi(s)ds\geq\delta
I_{n},\quad\forall t\geq0,
\]
which is equivalent to the previous formulation if $\dot{\varphi}(t)>0$
for all $t\geq0$, and also provides an additional degree of freedom
if $\dot{\varphi}(t)=0$ is allowed for some instants or intervals
of time.
\end{rem}

\section{Simulation Results}
\begin{figure}[ht]
	\centering
	\includegraphics[width=1\linewidth]{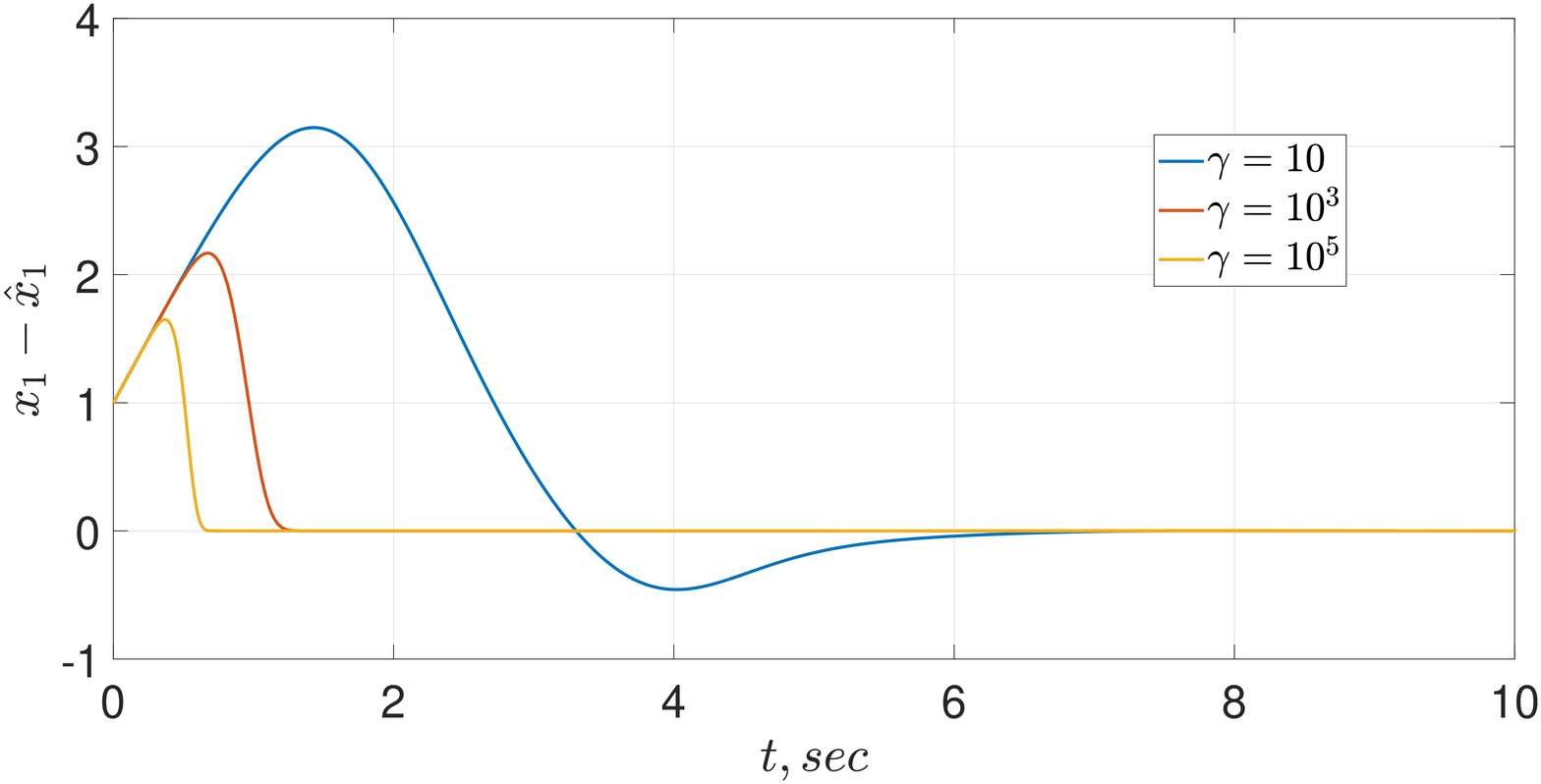}
	\caption{Error transients $x_1(t)-\hat{x}_1(t)$ for diffrerent $\gamma$ and case {\bf C1}}
	\label{fig_1}
\end{figure}

\begin{figure}[ht]
	\centering
	\includegraphics[width=1\linewidth]{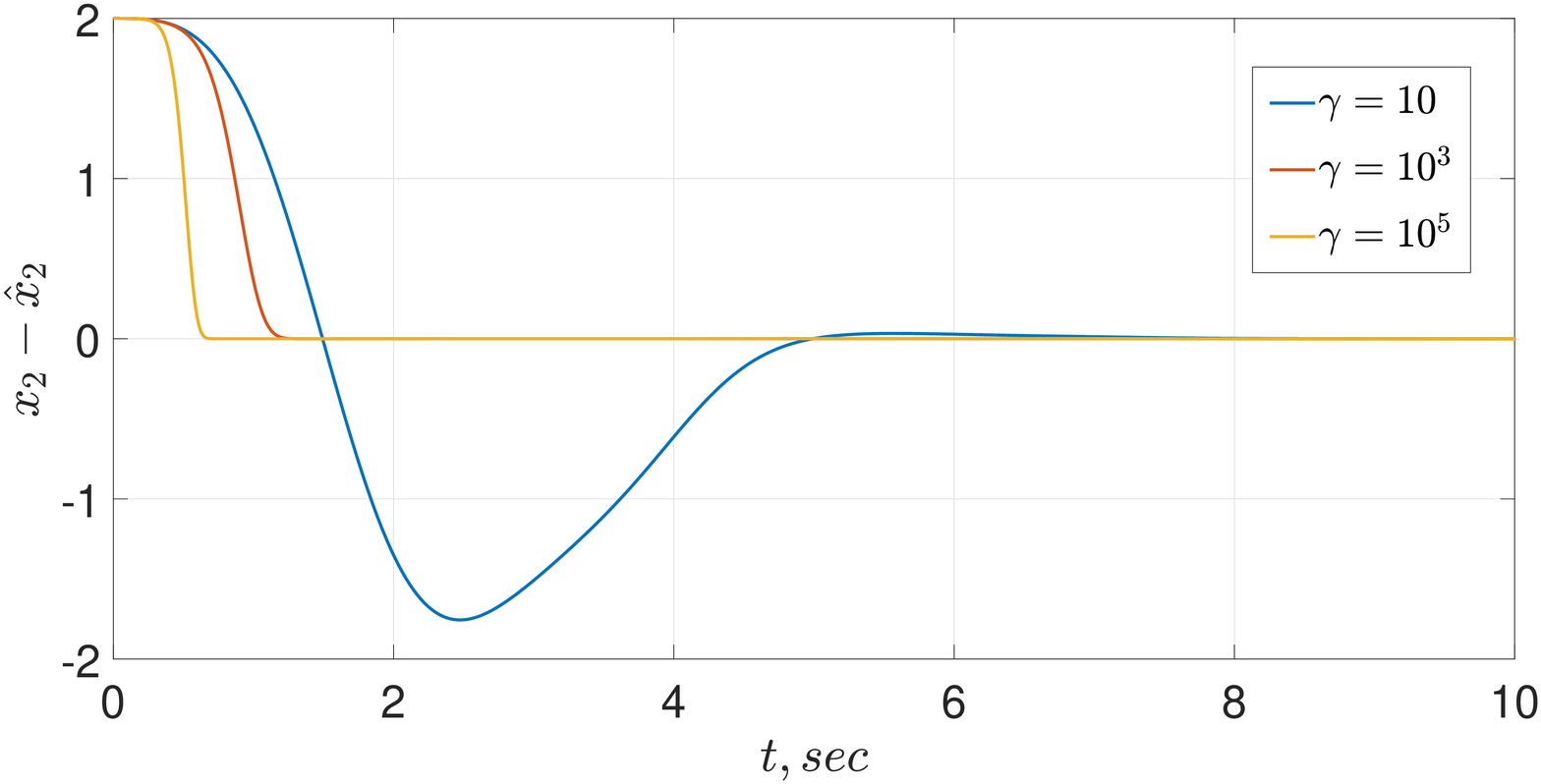}
	\caption{Error transients $x_2(t)-\hat{x}_2(t)$ for diffrerent $\gamma$ and case {\bf C1}}
	\label{fig_2}
\end{figure}

\begin{figure}[ht]
	\centering
	\includegraphics[width=1\linewidth]{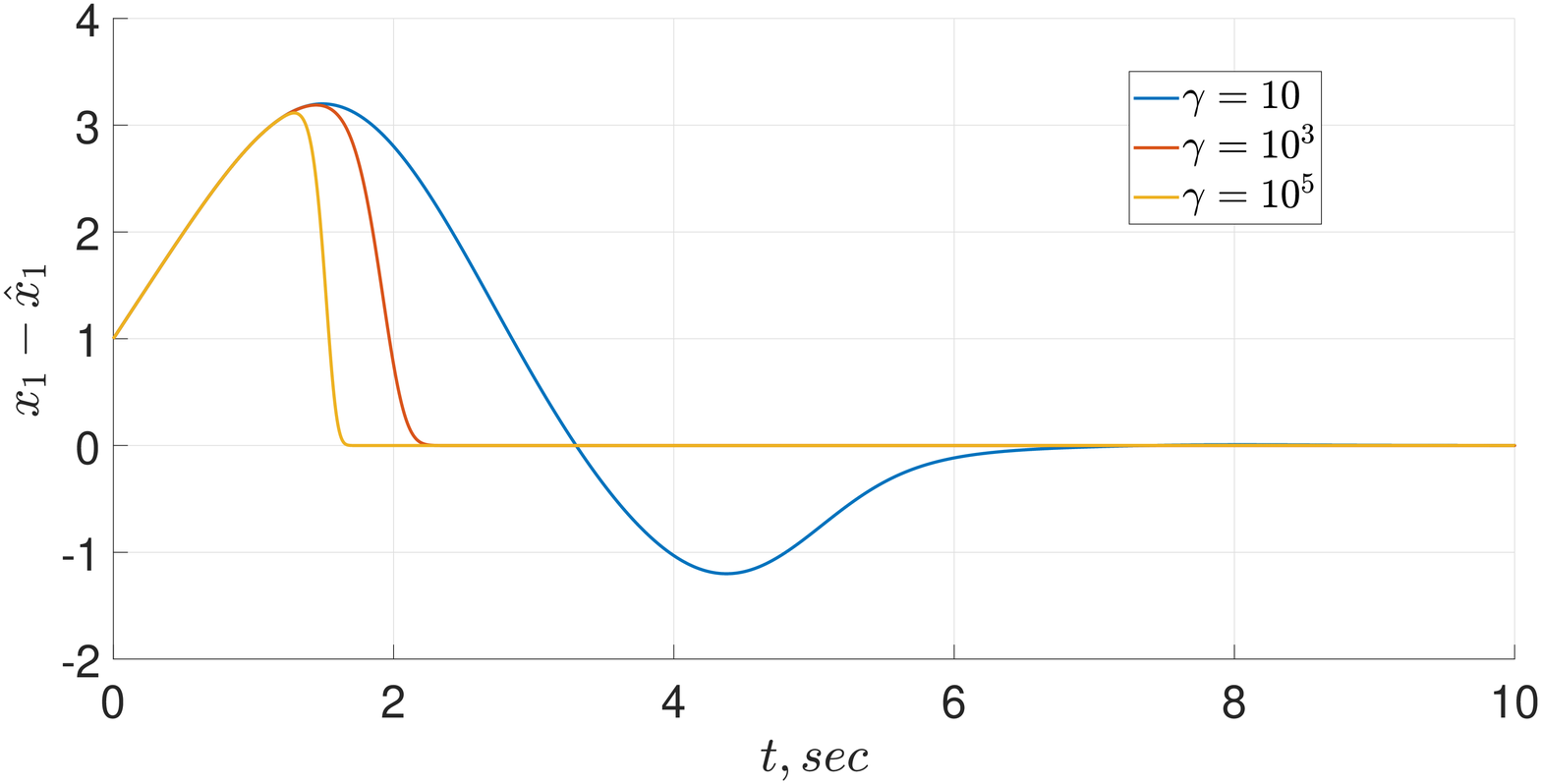}
	\caption{Error transients $x_1(t)-\hat{x}_1(t)$ for diffrerent $\gamma$ and case {\bf C2}}
	\label{fig_3}
\end{figure}

\begin{figure}[ht]
	\centering
	\includegraphics[width=1\linewidth]{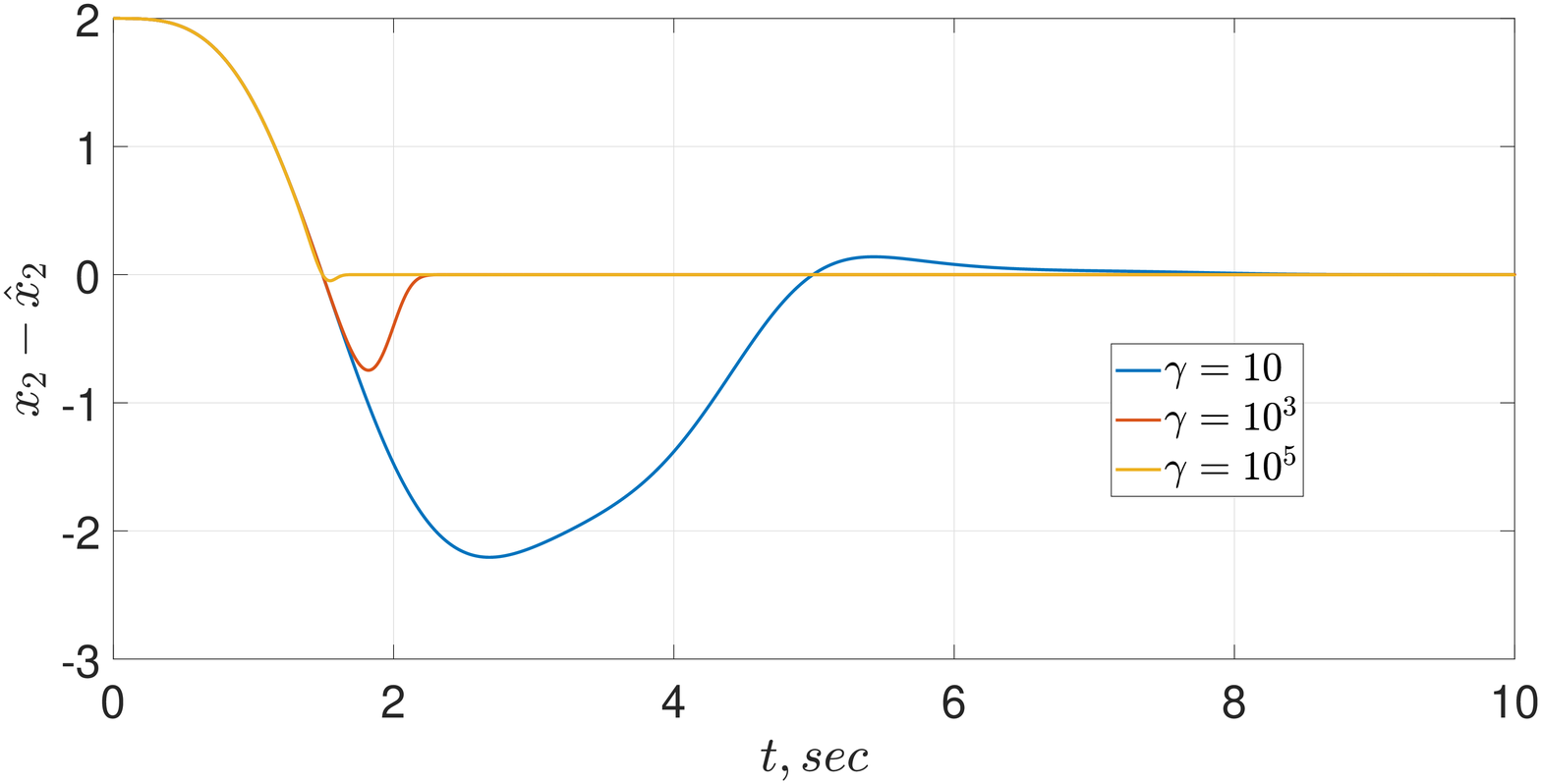}
	\caption{Error transients $x_2(t)-\hat{x}_2(t)$ for diffrerent $\gamma$ and case {\bf C2}}
	\label{fig_4}
\end{figure}

\begin{figure}[ht]
	\centering
	\includegraphics[width=1\linewidth]{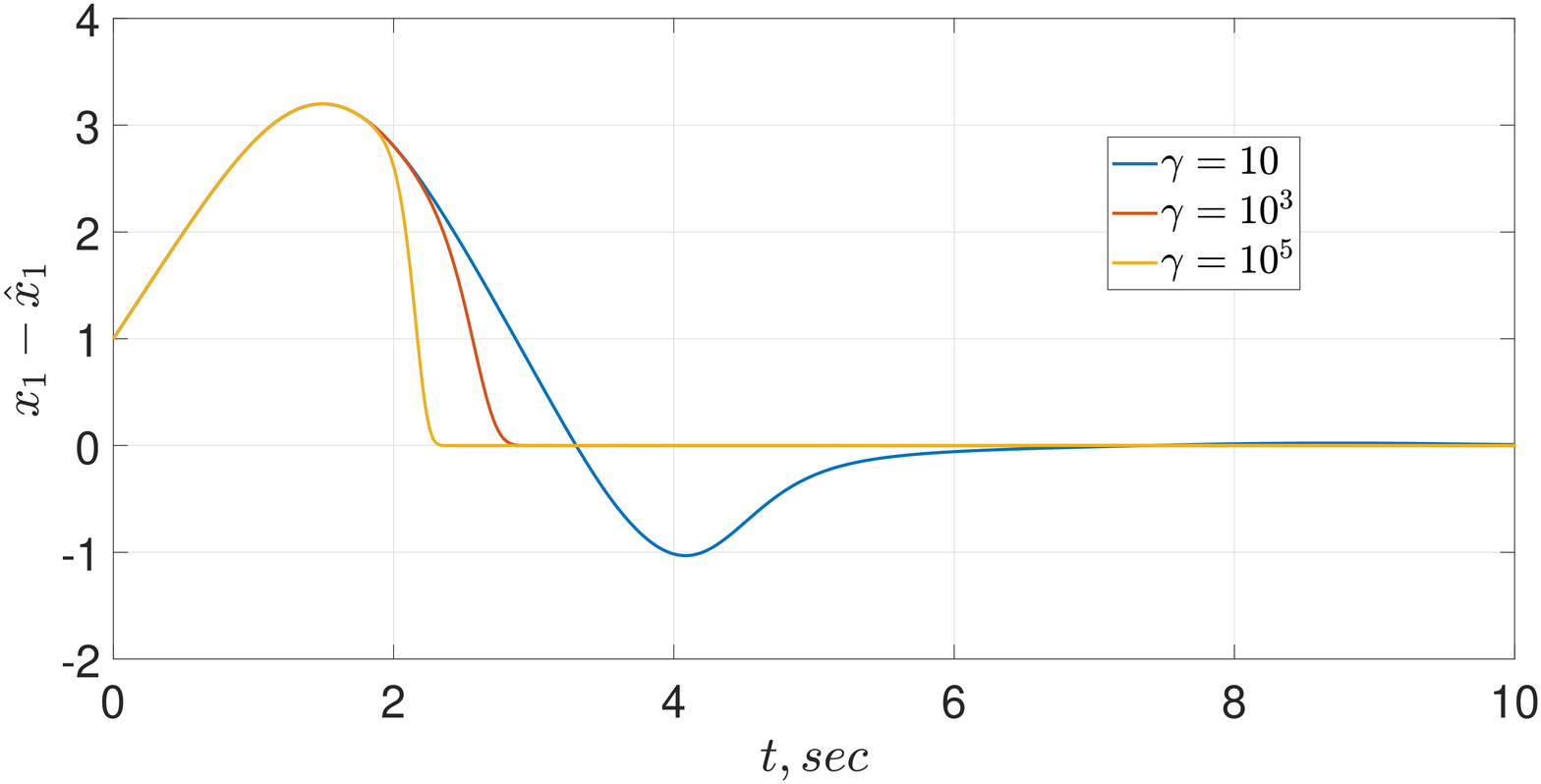}
	\caption{Error transients $x_1(t)-\hat{x}_1(t)$ for diffrerent $\gamma$ and case {\bf C3}}
	\label{fig_5}
\end{figure}

\begin{figure}[ht]
	\centering
	\includegraphics[width=1\linewidth]{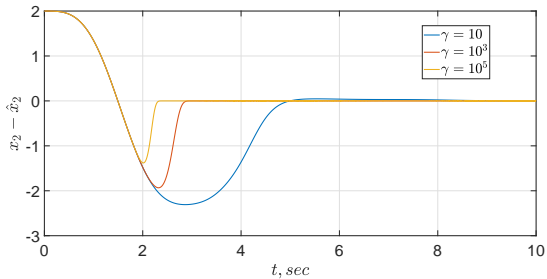}
	\caption{Error transients $x_2(t)-\hat{x}_2(t)$ for diffrerent $\gamma$ and case {\bf C3}}
	\label{fig_6}
\end{figure}

\lab{sec2}
%
Consider the LTV system \eqref{sys} with $m=q=1$, $n=2$ and
\begalis{
A &=
\begmat{
0 & 1\\
-\sin^2(t) & 0
},\; B=
\begmat{
0\\1
},\;C=\begmat{
	1\\0
}
}
For the estimation of $\theta$ we use the DREM approach \cite{ARAetaltac17} with $\Gamma=\gamma I_2$. We consider three cases:
\begenu[{\bf C1}]
	\item $\varphi(t)=t$ (Fig. \ref{fig_1} and Fig. \ref{fig_2});
	\item $\varphi(t)=\varphi(t-\tau)$, $\tau=1$ (Fig. \ref{fig_3} and Fig. \ref{fig_4});
	\item $\varphi(t)=\varphi(t-\tau)$, $\tau=1+0.9\sin(t)$ (Fig. \ref{fig_5} and Fig. \ref{fig_6});
\endenu

\section*{Acknowledgements}
%
This work was supported by the Ministry of Science and Higher Education of Russian Federation, passport of goszadanie no. 2019-0898, and by Government of Russian Federation (Grant 08-08).

%

\bibliographystyle{plain}        
\bibliography{gpebo_krstic_v4}           



\end{document}